# Origins Space Telescope: From First Light to Life

## ESA Voyage 2050 White Paper

M. C. Wiedner[1], S. Aalto[2], J. Birkby[3], D. Burgarella[4], P. Caselli[5], V. Charmandaris[6], A. Cooray[7], E. De Beck[2], J.-M. Desert[8], M. Gerin[1], J. Goicoechea[9], M. Griffin[10], P. Hartogh[11], F. Helmich[12], M. Hogerheijde[13], L. Hunt[14], A. Karska[15], Q. Kral[16], D. Leisawitz[17], G. Melnick[18], M. Meixner[19], M. Mikako[10], Ch. Pearson[20], D. Rigopoulou[21], T. Roellig[22], I. Sakon[23], J. Staguhn[17]

[1]Observatoire de Paris, PSL University, Sorbonne Université, CNRS, LERMA, Paris, France
[2]Chalmers University of Technology, Gothenburg, Sweden
[3]University of Amsterdam, Amsterdam, The Netherlands
[4]Laboratoire d'Astrophysique de Marseille, Marseille, France
[5]MPI for for Extraterrestrial Physics, Garching, Germany
[6]University of Crete, Crete, Greece
[7]UoC at Irvine, Irvine, USA
[8]University of Amsterdam, Amsterdam, The Netherlands
[9]CSIC, Madrid, Spain
[10]Cardiff University, Cardiff, UK
[11]MPI for Solar System Research, Goettingen, Germany
[12]SRON & Univ of Groningen, Groningen, The Netherlands
[13]Leiden University, Leiden, The Netherlands
[14]Osbservatorio Astrofisico di Arcetri, Florence, Italy
[15]Nicolaus Copernicus University, Torun, Poland
[16]LESIA, Observatoire de Paris, Université PSL, CNRS, Sorbonne Université, Univ. Paris Diderot, Sorbonne Paris Cité
[17]NASA/GSFC, Greenbelt, USA
[18]Center for Astrophysics | Harvard & Smithsonian, Cambridge, USA
[19]Space Telescope Science Institute, NASA/GSFC, Baltimore/ Greenbelt, USA
[20]Rutherford Appleton Laboratory, Didcot, UK
[21]University of Oxford, Oxford, UK
[22]NASA/Ames, Ames, USA
[23]University of Tokyo, Tokyo, Japan

## Abstract

The Origins Space Telescope (*Origins*) is one of four science and technology definition studies selected by National Aeronautics and Space Administration (NASA) in preparation of the 2020 Astronomy and Astrophysics Decadal survey in the US. *Origins* will trace the history of our origins from the time dust and heavy elements permanently altered the cosmic landscape to present-day life. It is designed to answer three major science questions: How do galaxies form stars, make metals, and grow their central supermassive black holes from reionization? How do the conditions for habitability develop during the process of planet formation? Do planets orbiting M-dwarf stars support life? *Origins* operates at mid- to far-infrared wavelengths from ~2.8 to 588 μm, is more than 1000 times more sensitive than prior far-IR missions due to its cold (~4.5 K) aperture and state-of-the-art instruments.

# 1 Executive Summary

**The Origins Space Telescope (*Origins*) will trace the history of our origins from the time dust and heavy elements permanently altered the cosmic landscape to present-day life. *Origins* operates at mid- to far-infrared wavelengths from ~2.8 to 588 µm, is more than 1000 times more sensitive than prior far-IR missions due to its cold (~4.5 K) aperture and state-of-the-art instruments.**

*Origins* investigates the creation and dispersal of elements essential to life, the formation of planetary systems, the transport of water to habitable worlds, and the atmospheres of exoplanets around nearby M-dwarfs to identify potentially habitable worlds. These science themes are motivated by their profound significance, as well as expected advances from, and limitations of, current and next-generation observatories (JWST, WFIRST, ALMA, and LSST). The nine key *Origins* scientific objectives (Table 1) address NASA's three major astrophysics science goals: *How does the universe work?, How did we get here,* and *Are we alone?* These nine aims also drive the instrumental requirements summarized in Table 2. The *Origins* design is powerful and versatile, and the infrared radiation it detects is information-rich. *Origins* will enable astronomers in the 2030s to ask new questions not yet imagined, and provide a far-infrared window complementary to planned, next-generation observatories (e.g., Athena, LISA, and ground-based ELTs).

**Table 1: Scientific objectives for the *Origins* Space Telescope**

| **NASA Goal** | **How does the Universe work?** | **How did we get here?** | **Are we alone?** |
|---|---|---|---|
| ***Origins* Science Goals** | 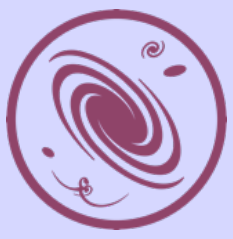 **How do galaxies form stars, make metals, and grow their central supermassive black holes from reionization to today?** | 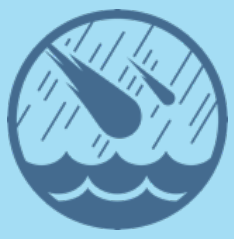 **How do the conditions for habitability develop during the process of planet formation?** | 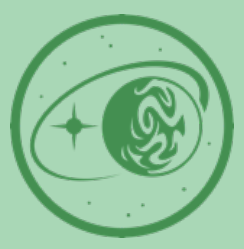 **Do planets orbiting M-dwarf stars support life?** |
| ***Origins* Scientific Capabilities** | *Origins* will spectroscopically 3D map wide extragalactic fields to simultaneously measure properties of growing supermassive black holes and their galaxy hosts across cosmic time. | With sensitive, high-resolution spectroscopy, *Origins* maps the water trail from protoplanetary disks to habitable worlds. | By obtaining precise mid-infrared transmission and emission spectra, *Origins* will assess the habitability of nearby exoplanets and search for signs of life. |
| ***Origins* Scientific Objectives** | 1) How does the relative growth of stars and supermassive black holes in galaxies evolve with time? 2) How do galaxies make metals, dust, and organic molecules? 3) How do the relative energetics from supernovae and quasars influence the interstellar medium of galaxies? | 1) What role does water play in the formation and evolution of habitable planets? 2) How and when do planets form? 3) How were water and life's ingredients delivered to Earth and to exoplanets? | 1) What fraction of terrestrial planets around K- and M-dwarf stars has tenuous, clear, or cloudy atmospheres? 2) What fraction of terrestrial M-dwarf planets is temperate? 3) What types of temperate, terrestrial, M-dwarf planets support life? |

The Origins Space Telescope (*Origins*) is one of four science and technology definition studies selected by NASA in preparation of the 2020 Astronomy and Astrophysics Decadal survey in the US. The study team included non-voting international representatives from ESA, JAXA, individual European countries, and Canada that contributed to the scientific and technical definition of Origins. A largely European team under French/CNES leadership designed the HEterodyne Receiver for Origins (HERO), which is one of the two upscope instruments.
The full Origins Mission Concept Study Report can be found at:
https://asd.gsfc.nasa.gov/firs/docs/
NASA has submitted four mission studies (LUVOIR, HabEX, Origins, and Lynx) to the Decadal survey and a prioritization is expected in 2021. If selected, Origins will begin Phase A in 2025 with a schedule that calls for a launch around 2035. This paper was submitted as a White Paper to the ESA Voyage 2050 call and describes the Origins Space Telescope to encourage ESA participation.

## 2. Key Science Goals and Objectives

### 2.1. How do galaxies form stars, make metals, and grow their central supermassive black holes from reionization to today?

*Origins* is designed to answer these fundamental questions in galaxy formation and evolution through wide area spectral mapping surveys in the far- infrared (FIR) wavelengths. *Origins* is capable of carrying out 3D infrared spectral mapping surveys resulting in spectroscopic data on millions of galaxies spanning the redshift range of z=0 to z > 6. These statistics are at a level comparable to the Sloan Digital Sky Survey (SDSS), but will be attained by *Origins* in a 2000 hours survey, instead of ~5 years from the ground in the optical that took to complete SDSS. Moreover, the FIR regime probes highly obscured environments that are more prevalent at higher redshift.

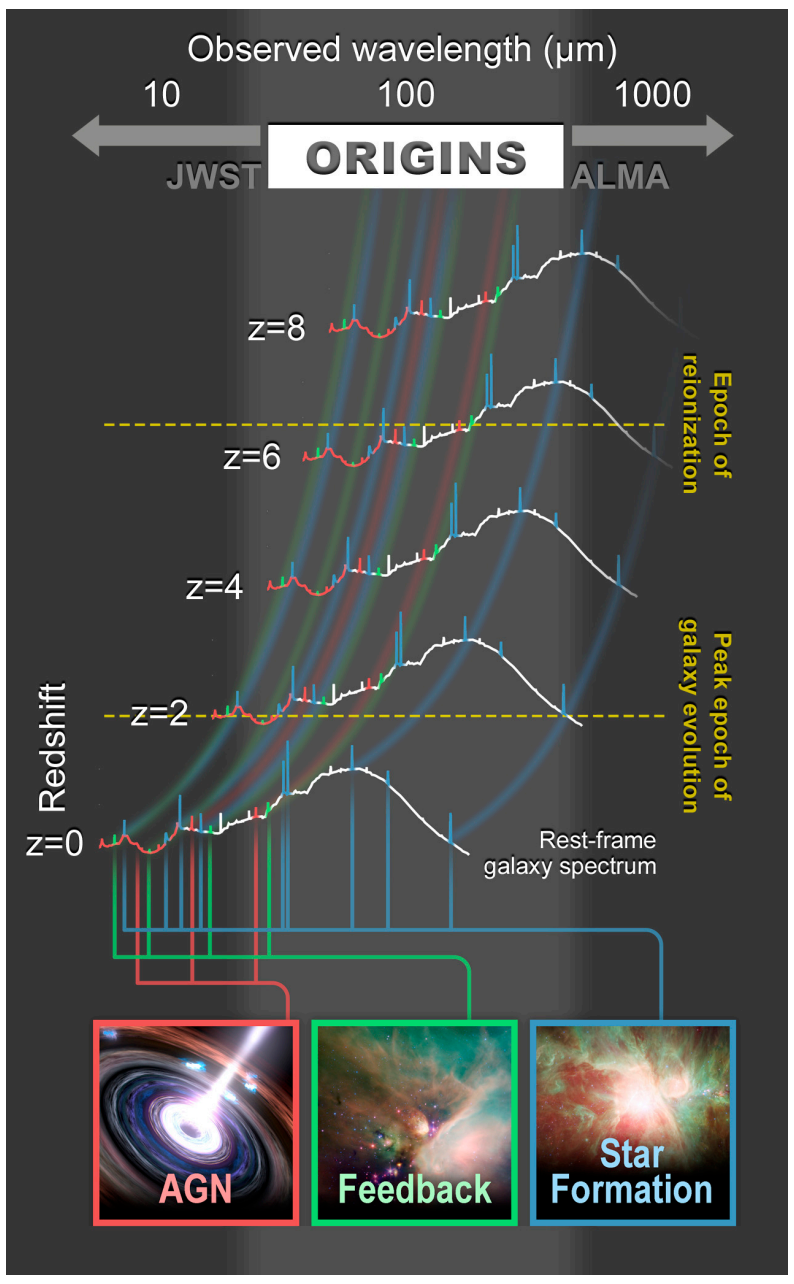

**Figure 1**. ***Origins*** **complements JWST mid-IR and ALMA sub-mm/mm-wave capabilities.** *Origins* will provide spectral line diagnostics indicative of AGN (red), star formation (blue), and feedback (green) over a wide range in redshifts, filling in a largely untapped region of wavelength and discovery space between JWST and ALMA.

A complete understanding of the astrophysical processes responsible for the formation and evolution of galaxies is one of the key scientific goals of modern-day astrophysics. While we have made significant strides, there are still huge gaps in our understanding of galaxy formation and evolution, especially the detailed astrophysical processes that grew and shaped galaxies over cosmic time. In particular, most of the accretion history of the Universe, both for star formation and for SMBH growth, took place in highly obscured environments. The FIR wavelength regime traces physical processes even in such extreme conditions, whose importance grows with increasing redshift. While small targeted surveys are capable of solving some key problems, uncertainties related to our models of galaxy formation are still strongly tied to small number statistics of galaxies at high redshifts and biases coming from galaxy selections at various wavelengths that are either sensitive to older stellar populations, such as in the near-IR, or active galactic nuclei (AGN) activity, such as in the X-rays. At far-IR wavelengths, spectral lines trace all key ingredients of galaxies providing multifaceted probe of internal processes in play in galaxies (Figure 1). *Origins*' wavelength range will not be explored by JWST, which is poised to provide the most detailed look yet at the distant universe. Furthermore, with a sensitivity that is a factor of 1000 improvement over *Spitzer* and *Herschel*, *Origins* capability moves beyond simply detecting rare "tip-of-the-iceberg" dusty, starbursting galaxies above the stellar mass vs. star-formation rate main sequence to studying dust, gas and AGN in the dominant galaxy populations. Finally, with 3D spectral mapping surveys, *Origins* overcomes issues related to source confusion that impacted previous continuum mapping surveys with *Herschel*.

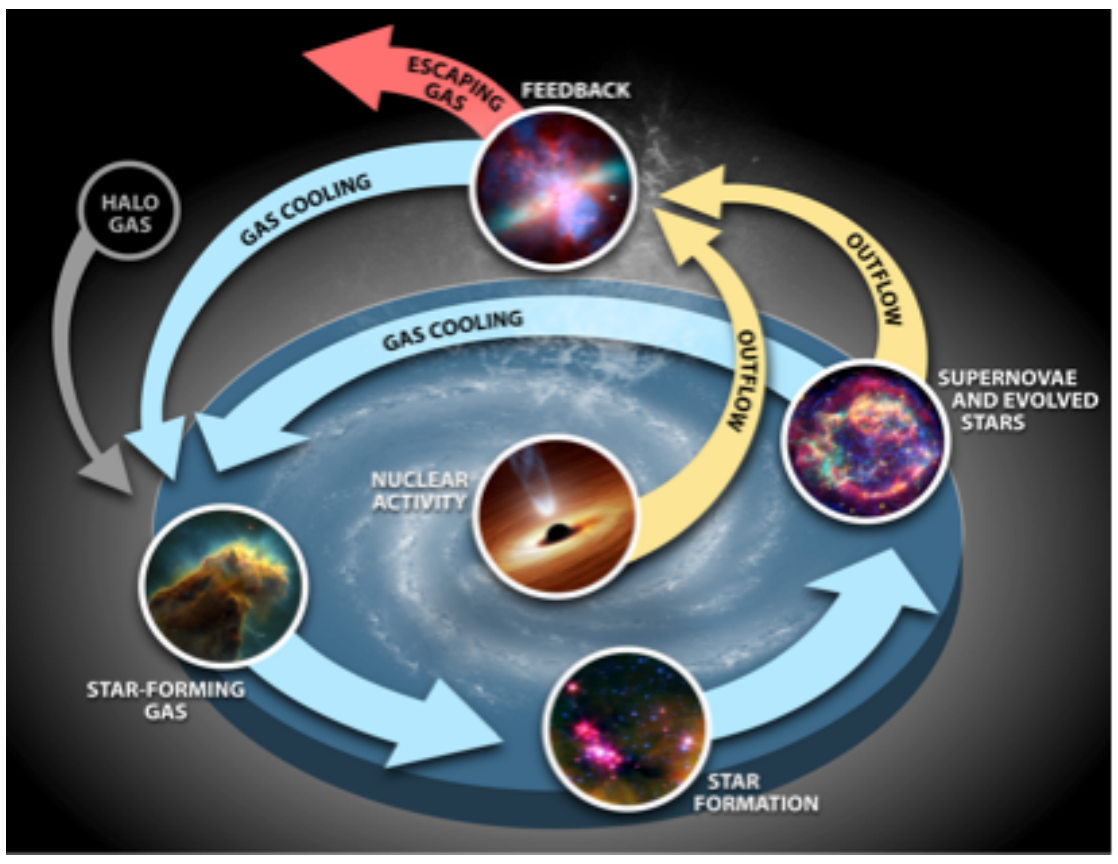

**Figure 2: *Origins* studies the baryon cycle in galaxies.** Energetic processes that shape galaxies and the circumgalactic medium together define the galactic ecosystem. Through its ability to measure the energetics and dynamics of the atomic and molecular gas and dust in and around galaxies that are actively star-forming or have AGN activity, *Origins* can probe nearly all aspects of the galactic ecosystem: star formation and AGN growth; stellar death; AGN- and starburst-driven outflows; and gas cooling along with accretion. These measurements will provide a complete picture of the lifecycle of galaxies.

**How do the stars and supermassive black holes in galaxies evolve with time?** *Origins* allows us to peer through the obscuring dust, probe the physics of star-formation through atomic and molecular gas, study the buildup of metals from dying stars, and establish the role of supermassive black holes (SMBH) as they accrete and drive energetic outflows into the surrounding interstellar medium (Figure 2). Key spectral signatures from the physical processes that sculpt galaxies are prominent in the infrared, where emission and absorption lines trace complex molecules, small and large dust grains, and atoms that are sensitive to changes in ionization and density (Pope et al. 2019).

**How do galaxies make metals, dust, and organic molecules?** Galaxies are the metal factories of the Universe, and *Origins* studies how metals and dust are made and dispersed throughout the cosmic web over the past 12 billion years. Sensitive metallicity indicators in the infrared can be used to track the growth history of elements via nucleosynthesis, even in the densest optically-obscured regions inside young galaxies at high redshift (metals: Smith et al. 2019; dust: Sadavoy et al. 2019).

**How do the relative energetics from supernovae and quasars influence the interstellar medium of galaxies?** Galaxies are made of billions of stars, yet star formation is extremely inefficient on all scales, from single molecular clouds to galaxy clusters. Because of its power to penetrate obscuration, *Origins* can study the role of feedback processes at play in galaxies over a wide range of environments and redshifts (Figure 1). *Origins* can study the processes that drive powerful outflows and map the demographics of galactic feedback (Bolatto et al. 2019 White Paper).

## 2.2 How do the conditions for habitability develop during the process of planet formation?

Water is essential for all life on our planet. Water provides the liquid medium for life's chemistry, while also playing an essential biochemical role. The formation story of water begins with the dynamical events, such as supernova explosions that gather and compress gas in the ISM and dust creating a latticework of filamentary clouds. It is within these dense clouds that stars and planets are born. Based on decades of study, we also know that water molecules formed as ice before stars are born in these dense clouds.

The Trail of Water begins as primordial interstellar material is provided to the young disk that will go on to form planets within tens of million years. A rotating collapsing cloud of gas and dust forms a young protostar surrounded by a disk that accretes material from a surrounding envelope. Within these proto-planetary disks, pebble-sized particles self-assemble under gravity to eventually form Earth-like worlds. Thanks to *Herschel* and *Spitzer*, and now with Atacama Large Millimeter Array (ALMA), we are able to construct the story of how these pre-planetary pebbles form and incorporate key molecules, such as water as ice from the natal cloud. However, without the ability to directly observe water in the forming disk, which *Origins* uniquely will provide, we cannot fully investigate this process.

### What role does water play in the formation and evolution of habitable planets?

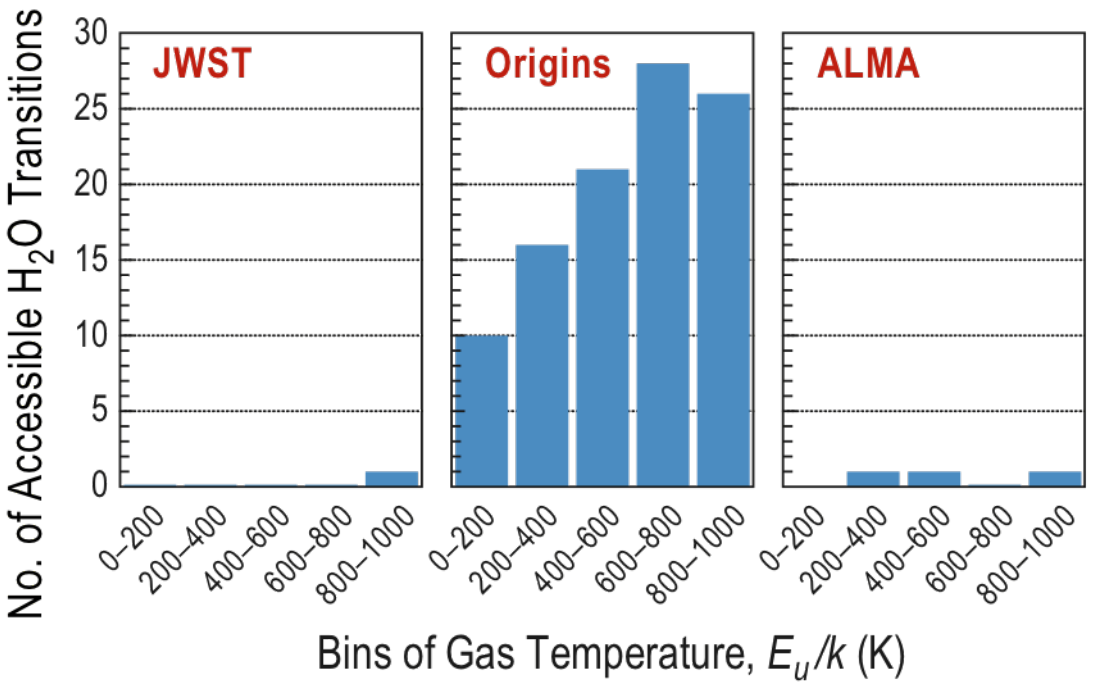

**Figure 3: Origins is capable of studying more than 100 transitions of water vapor, compared to one and three with JWST and ALMA, respectively.** This plot shows the number of $H_2{}^{16}O$ transitions observable by JWST, *Origins*, and ALMA as a function of the gas temperature, with energies above the ground state below 1000 K. ALMA is limited by atmospheric absorption in its ability to observe water lines from the Galaxy.

The broad wavelength coverage offered by *Origins* includes a large number of highly useful water vapor transitions unavailable to any other telescope, including ALMA or JWST. In fact, with Origins we can study nearly two orders of magnitude more water lines than we can with either ALMA or JWST. Just as importantly, these water lines cover an astounding range in temperature, from the snow line to the steam line in disks (Figure 3). With its unprecedented sensitivity to weak emission from less abundant forms of water, *Origins* provides the crucial measurement capability to understand water's role during key evolutionary phase that lasts a few hundred thousand years and ends when the young star ablates and dissipates the surrounding natal cloud (see Figure 4). Understanding the role of water in initial phase of formation from ISM to disks is the basis for the first scientific objective in this key mission design science program (Table 1).

## How and when do planets form?

During the subsequent phase that lasts a few million years when gas giants, such as Jupiter and Saturn, are born and the large Mars-sized embryos of Earth-like worlds are constructed, temperature plays a crucial role in determining what form of water will be found. If the environment is too hot, water will not exist as an ice but, instead, will be present as a vapor; if it is too cold, water will exist as an ice. Earth and the other terrestrial planets are constructed from coalescing solids. Over time, the newly-formed star gradually dissipates the gaseous disk, ending the phase of gas giant formation. The remaining disk is filled with rocky bodies both large (Mars-sized) and small (asteroids). It is during this time, over tens of millions of years, that terrestrial worlds such as our own formed. However, key elements of this picture remain uncertain since the ice-line may migrate as the disk evolves and planets are born. Using the HD 1-0 112 μm line as a tracer of the gas content, the second objective in this key mission design science program will establish the total gas mass in proto-planetary disks down to a mass limit of one Neptune.

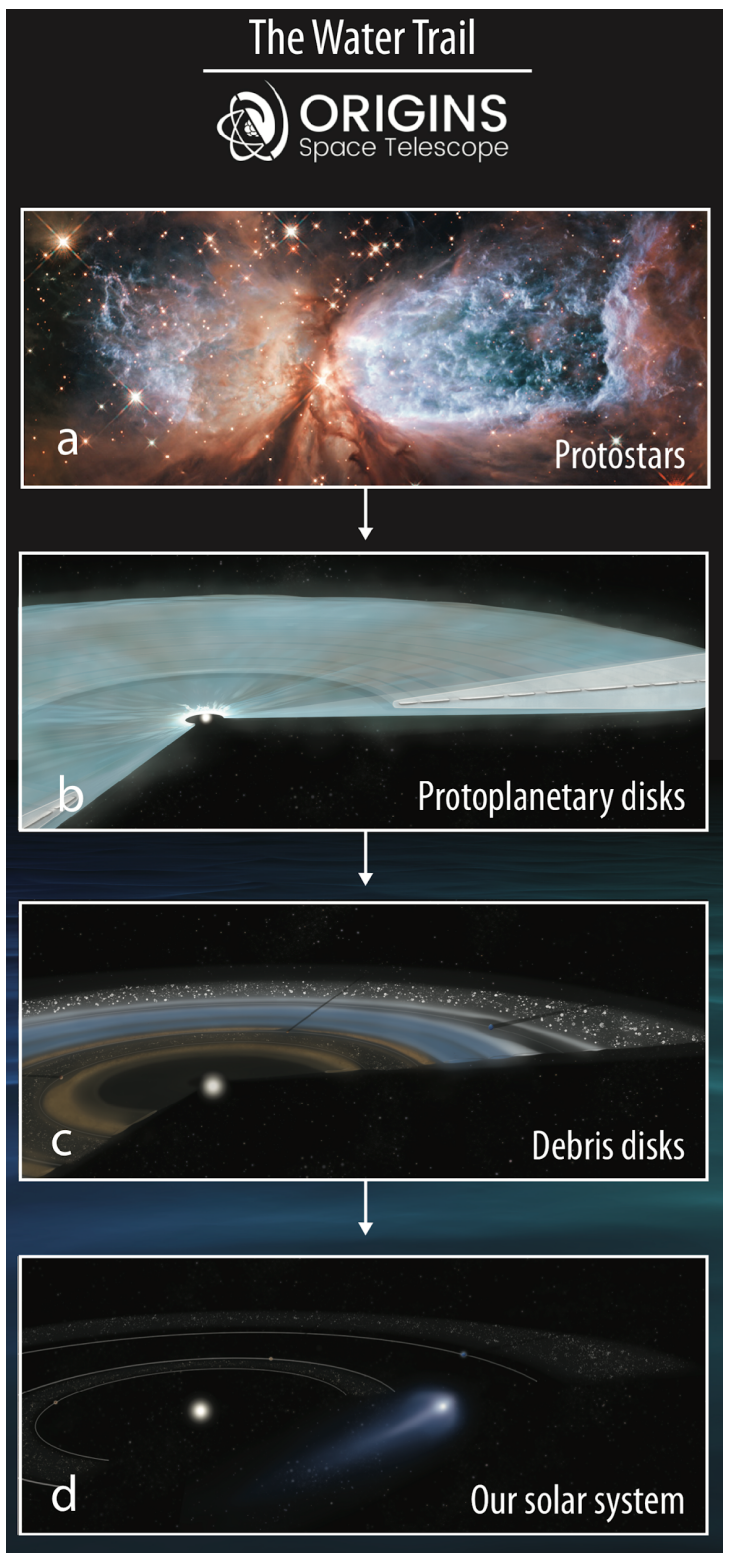

**Figure 4: *Origins* will trace water and gas during all phases of the formation of a planetary system.** The trail begins in the "pre-stellar" phase explored by Herschel, where a cloud of gas collapses (top) into a still- forming star surrounded by a disk nearly the size of our Solar System and a collapsing envelope of material (2nd from top). Over time, the envelope dissipates, leaving behind a young star and a disk with nascent planets (3rd from top), eventually leaving behind a new planetary system (bottom). *Origins* will excel at probing the protoplanetary and later phases.

## How were water and life's ingredients delivered to Earth and to exoplanets?

The dynamical interactions and the construction of rocky worlds by energetic impacts leads to a phase that generates significant "debris" from the collisions. The "ice-line" – the distance from the young star where water transitions from a gas to a solid – holds a prominent place in planet-formation, as it is believed that the Earth, and its precursor materials, formed inside our Solar System's ice-line. It is theorized that water was delivered to the early Earth via impacts from material that formed beyond the ice-line during this debris phase. While debris disks have been mapped using prior space telescopes, they did not provide the capability to determine whether the impactors carry water. Consistent with the picture of water delivery, there is one revealing piece of evidence that suggests the Earth received its water from somewhere quite cold. This evidence lies in the fraction of deuterium in water. Earth's water has an excess of deuterium (i.e., heavy water), and simple chemical principles inform us that this excess could only have been created when water is formed at a temperature of 10-20 K. In our Solar System, comets and asteroids all carry this signature and, in principle, we can then use the deuterium fraction to trace back the primary source of Earth's water to either the asteroid belt or larger distances where comets reside. Unfortunately, we currently have measurements toward only a handful of comets, all hinting at subtle D/H variations. By making measurements toward hundreds of comets, *Origins* will finally establish whether asteroids and/or comets were the source of Earth's water. The third science objective in this key mission design science program will target close to 100 comets during the 5-year lifetime to measure D/H (Table 1). The combination of these three objectives and their proposed measurements focused primarily on water and HD1-0 will transform our understanding of Earth's evolution as well as the mechanism by which habitable planets form and obtain the key life-enabling ingredient, water.

## 2.3 Do planets orbiting M-dwarf stars support life?

Humankind has long pondered the question, *Are we alone?* Only now are scientists and engineers designing instruments that are dedicated to answering this question. Our quest to search for life on extra-solar planets relies on our ability to measure the chemical composition of their atmospheres. *Origins* expands upon the legacy of *Hubble* and *Spitzer* – and soon JWST – with a mid-infrared instrument specifically designed for transmission and emission spectroscopy measurements. In its search for signs of life, *Origins* employs a multi-tiered strategy, beginning with a sample of planets with well-determined masses and radii that are transiting nearby K & M dwarfs, the most abundant stars in the Galaxy. With its broad, simultaneous wavelength coverage and unprecedented stability, *Origins* is uniquely capable of detecting atmospheric biosignatures (Figure 5).

**What fraction of terrestrial K- and M-dwarf planets has tenuous, clear, or cloudy atmospheres?** In the first tier of its exoplanet survey, *Origins* will obtain transmission spectra over the 2.8–20 μm wavelength range for temperate, terrestrial planets spanning a broad range of planet sizes, equilibrium temperatures, and orbital distances, in order to distinguish between tenuous, clear, and cloudy atmospheres. Because $CO_2$ absorption features are so large, this tier can include terrestrial planets orbiting stars from late-M to late-K, giving *Origins* a broader perspective in the search for life than JWST.

**What fraction of terrestrial M-dwarf planets is temperate?** For a subset of planets with the clearest atmospheres, *Origins* will measure their thermal emission to determine the temperature structure of their atmospheres. This measurement is critical to assessing climate because it yields an understanding of how incoming stellar and outgoing thermal radiation dictate the heating and cooling of the atmosphere. *Origins* can then determine whether these atmospheric conditions could support liquid water near the surface (Line et al. 2019).

**What types of temperate, terrestrial M-dwarf planets support life?** *Origins* will be the first observatory with the necessary spectroscopic precision to not only measure habitability indicators ($H_2O$ and $CO_2$), but also crucial biosignatures ($O_3$ coupled with $N_2O$ or $CH_4$), which are definitive fingerprints of life on habitable-zone planets. In this observational third tier, *Origins* will obtain additional transit observations for the highest-ranked targets to search for and detect biosignatures with high confidence. The wavelength range afforded by *Origins* will provide access to multiple spectral lines for each molecular species. This will increase the detection significance and prevent potential degeneracies due to overlapping features, thus averting false-positive scenarios. This framework robustly detects a variety of potentially habitable planet atmospheres, including the life-bearing Archaean Earth. The entire era of exoplanetary atmospheres, including the life-bearing science has shown that Nature's imagination trumps our own and *Origins'* broad wavelength coverage and precise measurements are guaranteed to give us views into the new and unexpected in the domain of life elsewhere in the Galaxy (Kataria et al. 2019).

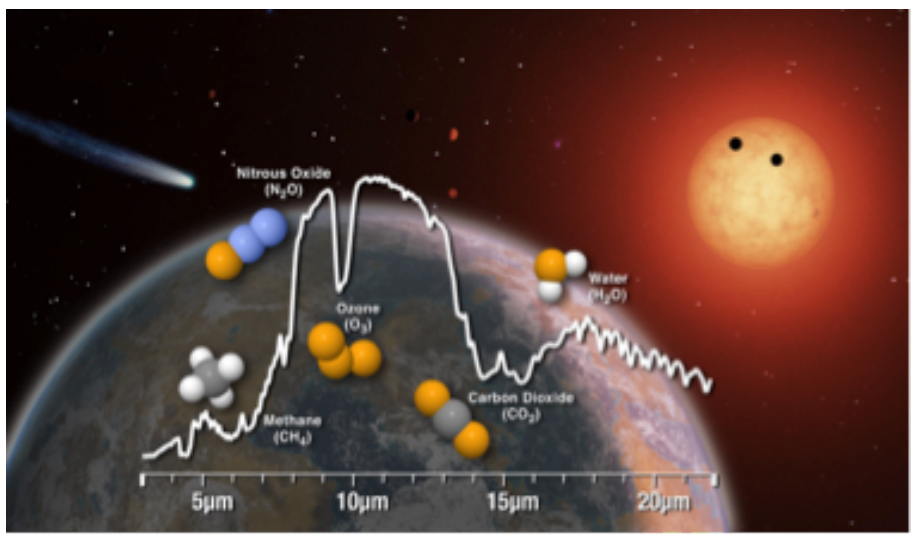

**Figure 5: *Origins* is designed to search for atmospheric biosignatures of exoplanets that transit K- & M-dwarf stars.** By leveraging the mid-infrared wavelength capabilities with a dedicated state-of-the-art instrument for exoplanet transit and eclipse studies, *Origins* will study the exoplanet atmospheres for gases that are the most important signatures of life.

## 2.4 Discovery science for Origins in its baseline configuration

With more than three orders of magnitude improvement in sensitivity over *Herschel* and access to a spectral range spanning nearly 8 octaves, *Origins* vastly expands discovery space available to the community. While the mission is designed to achieve a specific set of objectives, the science program is intended to be illustrative only. *Origins* is a community observatory, driven by science proposals selected through the usual peer-review process, as with existing NASA observatories.

Suggestions for the discovery science for Origins include:

- Detection of warm molecular hydrogen from reionization
- Mapping Galaxy outflows in the nearby universe
- Wide-field mapping of molecular hydrogen in local group dwarf galaxies
- Mapping Magnetic fields at galactic scales
- Follow-up and characterization of LISA and LIGO gravitational-wave sources and other time-domain sciences
- Time-domain sciences: proto-star variability as a probe of protoplanetary disk physics and stellar assembly
- Small bodies in the trans-Neptunian region: constraints on early solar system cometary source region evolution
- Origins studies of water ice in non-disk sources
- Studying magnetized, turbulent molecular cloud
- Below the surface: A deep dive into the environment and kinematics of low luminosity protostars
- Putting the Solar System in context: the frequency of true Kuiper-belt analogues
- Giant planet atmospheres: templates for brown dwarfs and exoplanets

*Origins'* sensitivity exceeds that of its predecessor missions by a factor of 1000. Jumps of this magnitude are very rare in astronomy, and have always revolutionized our understanding of the Universe in unforeseen ways. Thus, it is essentially guaranteed that the most transformative discoveries of *Origins* are not even anticipated today.

## 2.5 Discovery science requiring HERO

The *Origins* baseline concept proposes three extremely powerful instruments: OSS, FIP, and MISC. However, these instruments are incapable of fully investigating the trail of water from the cold interstellar medium to planet forming disks and solar system objects.

### Early stages of the trail of water

With its heterodyne receiver, *Origins* in its upscope configuration will play a critical role in tracing the early path of water from the ISM into young circumstellar disks through its unique access to the lowest-energy rotational transitions of the water molecule and its isotopologues ($H_2{}^{18}O$, $H_2{}^{17}O$, HDO) at high spectral resolving power (up to $10^7$), and in synergy with JWST for tracing water ice through its infrared and far infrared bands. With the high sensitivity provided by its large, 5.9-m telescope and HERO's extremely high spectral resolution capabilities, *Origins* will be a transformational tool for following the path of water in the ISM.

While some interstellar water is known to be present in diffuse molecular gas and UV-irradiated photodissociation regions, the bulk of water is found in dense molecular clouds as ice mantles on cold (T ~ 10 K) dust grains with tiny traces of water vapor (three orders of magnitude less abundant than water ice). Via a host of observations (e.g., Whittet et al., 1983; Öberg et al., 2011; Boogert et al., 2015), it is now known that the water ice mantle first forms in pre-stellar cores. This is the water that is provided to the young disk and sets the stage for all that follows.

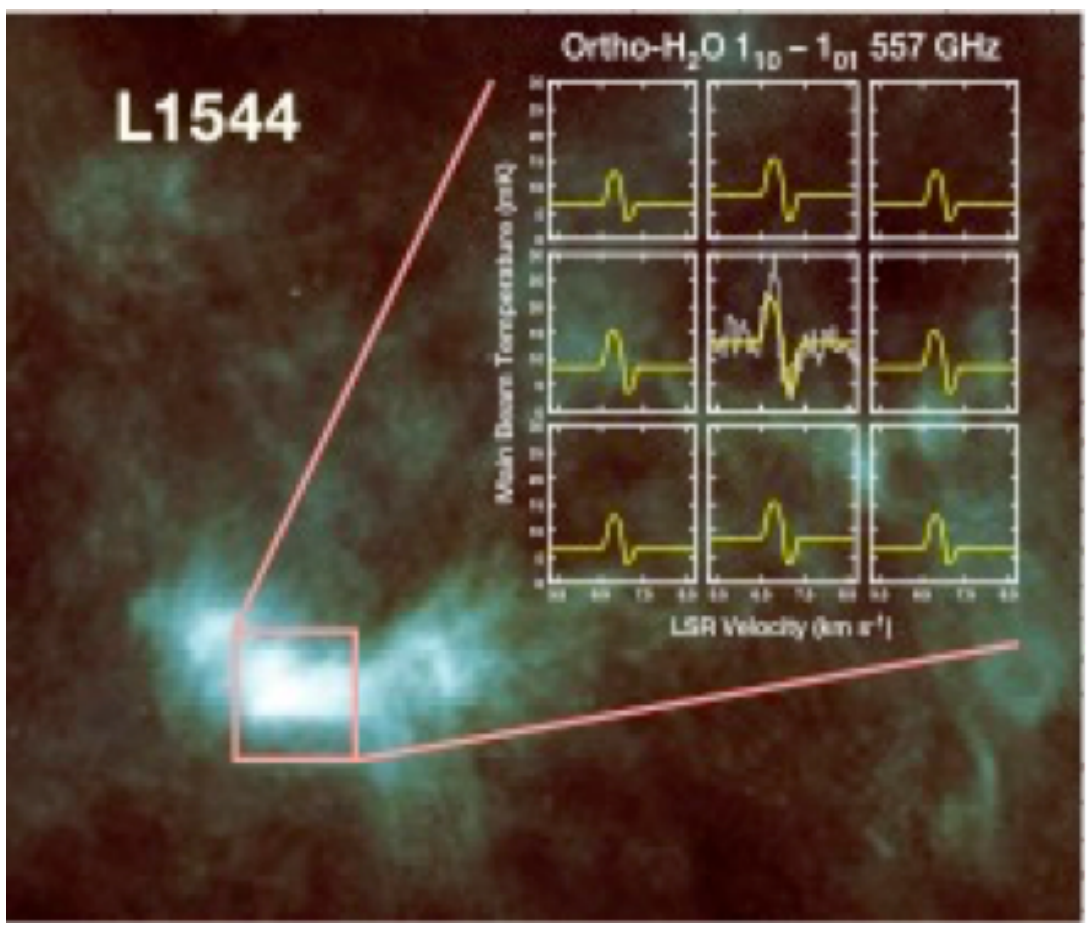

**Figure 6:** High spectral resolution observations reveal complex line shapes of the ground state water line that allow us to determine the dynamics of the inner cores. The L1544 prestellar core showing the dust continuum emission. The insert is the modeled line profiles of the $H_2O$ $1_{10}$-$1_{01}$ transition at 538 µm (557 GHz) in the 25" FWHM beam of Origins using the MOLLIE radiative transfer code (Keto et al., 2014). Each panel is separated by 25". The central panel includes the HIFI spectrum with a 40" beam. Note the higher continuum level and deeper absorption in the central panel, which shows the forming core.

State-of-the-art chemo-dynamical models of the prestellar core evolution that include water ice and cosmic ray-induced production of water vapor predict that, overall, for a typical prestellar core of 1 solar mass ($2x10^{33}$ g), the total mass of water vapor can be estimated to lie between 20 and $2x10^3$ Earth ocean mass, while the total mass including water ice would be up to a few millions Earth ocean mass. The spread in these estimates is not only due to the individual variation between cores related to their environment, but also to the lack of adequate observing facilities since Herschel.

There is only a single published observation of detection of water vapor in a starless core (Caselli et al., 2012) that highlighted the role of cosmic rays and provided definitive evidence for a gravitationally collapsing core. Because of their sensitivity to hydrogen densities higher than $10^7$ $cm^{-3}$, the ground state water line profiles uniquely reveal the dynamics of the inner regions of the collapsing core, enabling a clear and accurate test of star formation theories and an accurate measurement of the amount, radial distribution and infall speed of water that is delivered to the forming protostar/protoplanetary system (Keto, Rawlings, & Caselli, 2014, 2015). The main gas phase precursors, OH and $H_3O^+$, will be accessible to HERO, leading to a complete account of the chemical network of water. This will allow us to follow the formation of water during the dynamical evolution of starless cores, on their way of star and stellar system formation. Measurements of water vapor in different environments will also provide important clues on how physical parameters (e.g. the impinging radiation field, volume densities, dust and gas temperatures and turbulent content) affect the production of water and its accumulation on dust grains, the building blocks of pebbles and planets. This is crucial to put stringent constraints on our chemical-dynamical models. With its upgraded instrument suit *Origins* is uniquely suited to perform a deep survey of cores at different stages of evolution and in different star-forming environments.

HERO is a powerful instrument to zoom into the disk using line-tomography:

- The high spectral resolving power of 30 m/s allows much more detailed localization of the gas than OSS, especially for the cold regions in large disks, which have Keplerian velocities as low as a 1-2 km/s.
- Line tomography will be possible over the whole frequency range of HERO, including the lowest-lying ortho-water line at 538 µm, which traces the coldest gas (DE/k = 27K), and is significantly less effected by dust attenuation than the 179.5 micron line, giving access to deeper layers, even in massive disks.
- A single setting of HERO observes both the ortho-line at 538 µm, and the para-line at 269 µm, and includes the entire line profile with at least 100 resolution elements without the need for scanning.

## Final stages of the trail of water

HERO provides high spectral resolution observations of water in our solar system that allows us to understand in detail how water was transported to Earth and where it is found in our solar system. We can apply this knowledge to other planetary systems also to help our search for life.

Comets are one possible origin of water on Earth, where the D/H ratio is an important indicator of the likelihood that comets delivered water found on Earth (Hartogh et al, 2011a). Complementary to the large OSS comet survey, HERO carries out follow up observations of the brighter comets at very high spectral resolving power ($10^7$ ) in order to determine the origin of outgassing for different molecules (by analyzing the line profiles), to get a refined D/H ratio, to determine the excitation mechanisms of the gases, and to determine the gas coma structure. HERO contributes to questions about the origin of the solar system by providing isotopic ratios (e.g. D/H, $^{16}O/^{17}O$ and $^{16}O/^{18}O$) for a large number of comets and linking the age of these objects to other primitive bodies.

Water does not only exist on Earth, but also on other planets, such as Mars, which may have supported life. HERO observations constrain the water cycle (Shaposhnikov et al., 2019), hydrogen/oxygen chemistry and origin of the Martian atmosphere by very sensitive and highly spectrally-resolved observations of molecules and their isotopologues (Villanueva et al., 2015) and provide their vertical profiles from the pressure broadened line shapes. Furthermore it will provide upper limits for a large number of molecules so far not detected (e.g. HCl). HERO constrains the origin of water in the stratosphere of the gas and ice giants, determines the D/H ratio in hydrogen and water and the isotopic ratios of at least C, S and O with high precision. Moons of these planets also contain water. HERO is capable of monitoring the composition, physical conditions and variability of the Enceladus torus (Hartogh et al, 2011b), the water atmospheres of the Galilean satellites (including detection of plumes), Titan (Moreno et al.. 2012) and the dwarf-planet Ceres (Küppers et al. 2014).

## ***Origins*** as an Event Horizon Telescope station

The concept of using *Origins* to study black hole physics on event horizon scales was recently raised as a potential extension of the HERO science case, that shows much promise and merits further study. The unprecedented angular resolution resulting from the combination of *Origins* with existing ground-based submillimeter/millimeter telescope arrays would increase the number of spatially resolvable black holes by a factor of $10^6$, permit the study of these black holes across all cosmic history, and enable new tests of General Relativity by unveiling the photon ring substructure in the nearest black holes, see Figure 7. Expanding the HERO instrument to be an interferometric station will require small technology enhancements. (For more details: Origins Space Telescope Mission Concept Study Report, App D-16.)

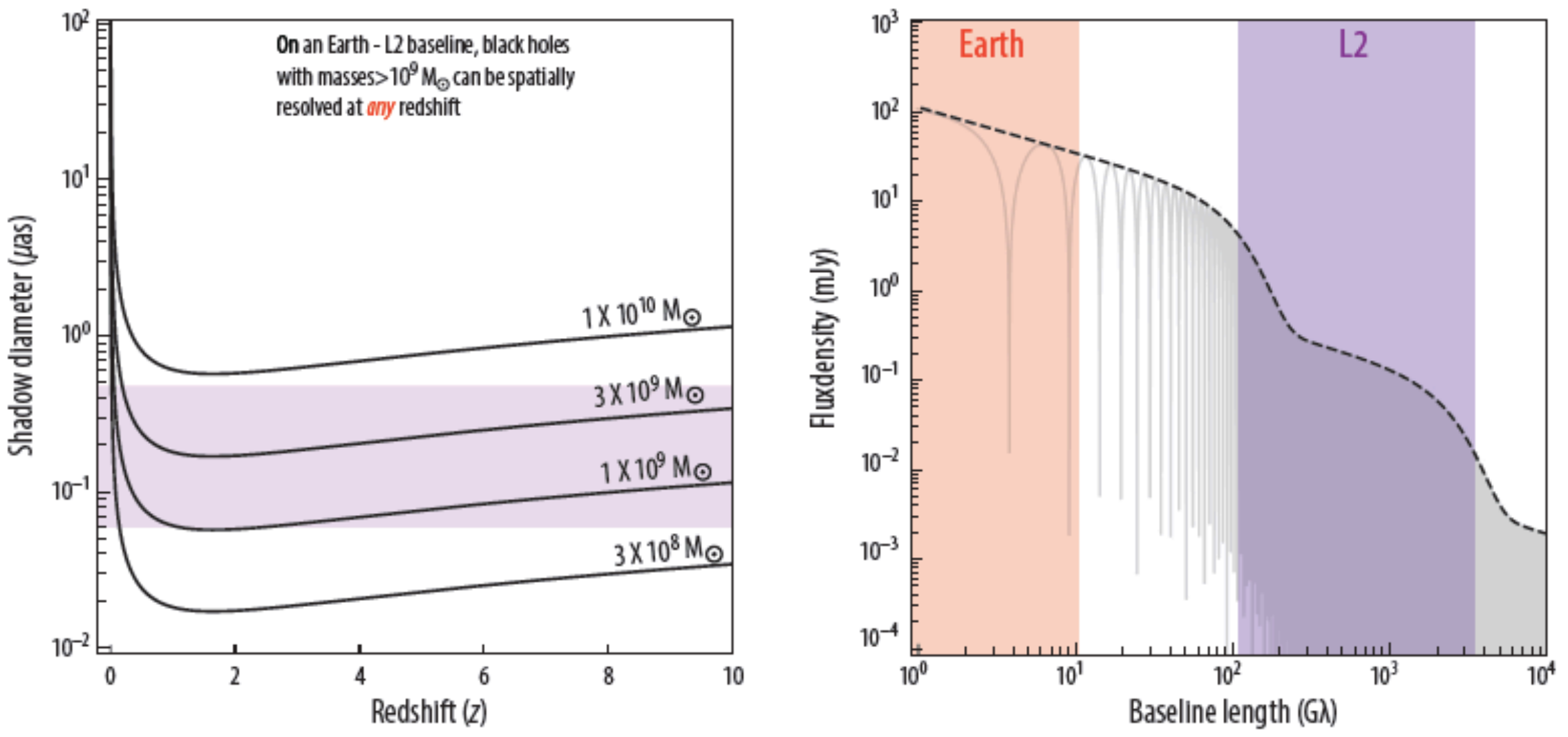

**Figure 7: Left: Black hole shadow diameter versus redshift for SMBHs. Right: Flux density as a function of interferometer baseline length. In both plots the parameters for the L2-Earth baseline is shaded in purple.**

## 2.6 Discovery science requiring MISC-upscope

### The rise of Metals

Origins/OSS will use sensitive metallicity indicators in the infrared to track the growth history of dust. At z > 5 galaxies are generally metal poor and bright rest-frame optical lines as well as the 3.3micron PAH feature that fall into the mid-IR are best suited to trace the dust. The 3.3. micron PAH feature is particularly interesting as it can be used to track metallicity as long as it is not destroyed by high UV radiation fields typically found in low metallicity environments. The MISC upscope spectrometer allows these observations. Observations of different mid-IR emission lines also carry information about the nature of photo-ionization in individual galaxies.

### Building the stellar masses of early galaxies (WFIRST + Origins/MISC)

Stars emit photons over the entire wavelength range, but the unreddened emission from stars in the early universe at 5 < z < 10 falls in the rest-frame UV, optical and near-IR (roughly 0.25 – 3.5 µm). This information is used to estimate the star formation rate (SFR). To get a picture of the history of stellar formation, we also need to measure the integral of this SFR(t), e.g., the stellar mass of galaxies. This translates into wavelengths that uniquely match OST/MISC's, i.e. 5 – 30 µm. The rest-frame UV spectrum will provide an access to young stars that are likely to be predominant at z > 5. We can collect this information from WFIRST-Deep and WFIRST-Wide surveys. But, if we want to perform a complete census, included potential older stars, we need the rest-frame optical and near-IR. To follow up WFIRST's objects at z > 5, after JWST lifetime, we need an instrument like MISC on OST. The ELT might bring some information, though, but mainly below 2.5µm. Using Origins/MISC to study the galaxies detected at 5 < z < 10 in WFIRST-Deep and WFIRST-Wide surveys will enable critical measurements of the star formation rate (SFR), stellar mass (M«), and dust attenuation. Two surveys using Origins/MISC WFI photometric and spectroscopic capabilities will provide unique data that broaden our understanding of the evolution of the mass function (cosmic mass assembly), star formation rate density, and average dust attenuation for a representative sample of galaxies at $5 < z < 10$. The WFI-L channel (9-28μm) is prioritized rather than WFI-S (5-9μm) in the course of the descope discussion by the STDT.

## 2.7 Origins: A Mission for the Astronomical Community

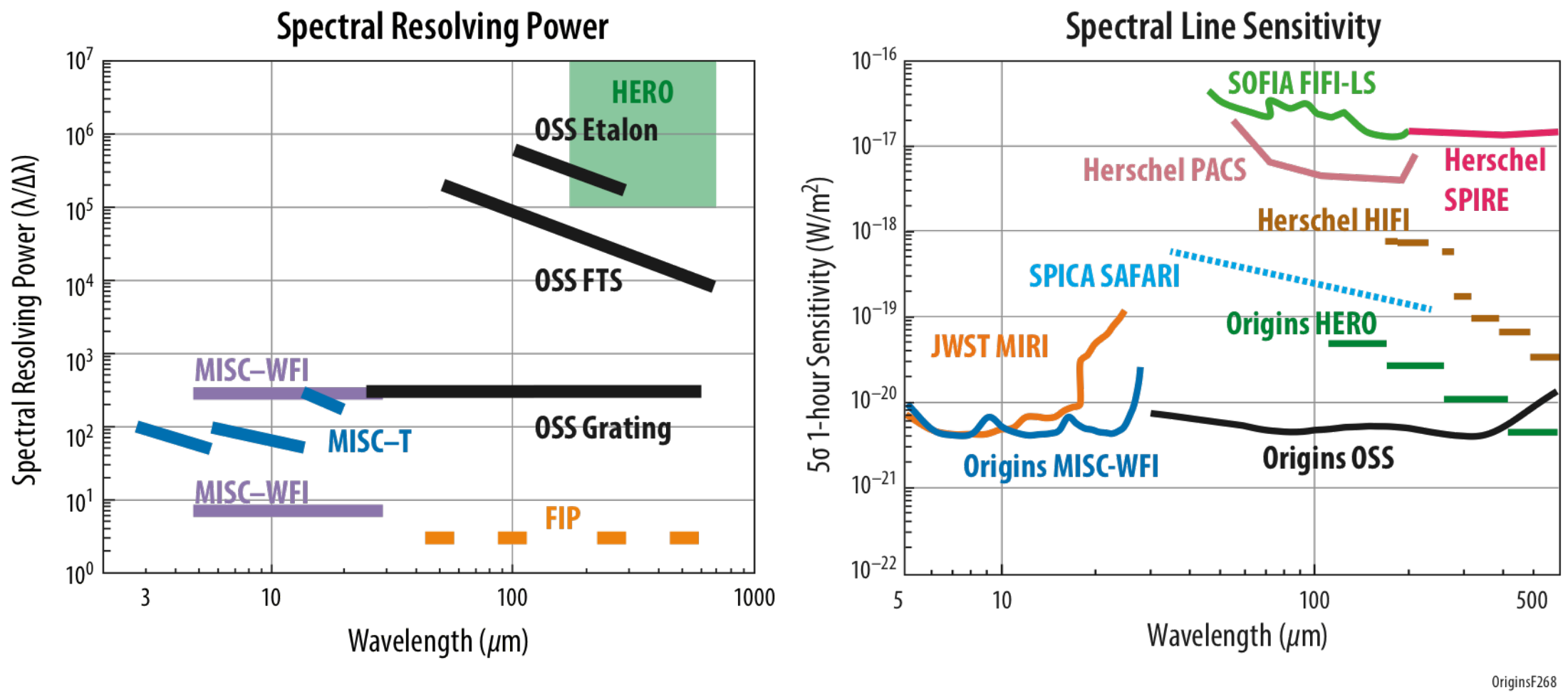

**Figure 8: *Origins* has a comprehensive set of three baseline instrument and two upscope instrument options that make it a very versatile and powerful instrument for key science questions as well unanticipated discoveries.** *Origins* surpasses all prior, current and planned mission by a large factor setting the stage for a new era of far-IR astronomy.

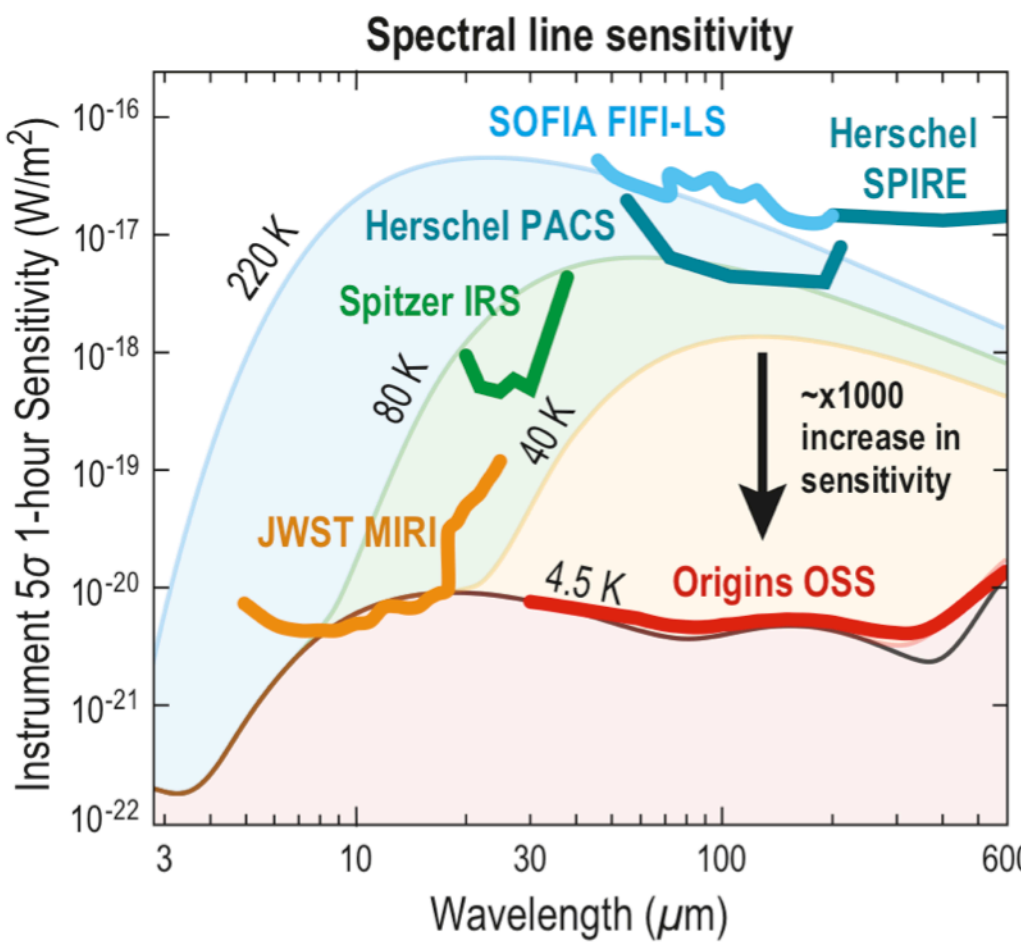

**Figure 9: *Origins* taps into a vast, unexplored scientific discovery space, defined by a three-orders-of-magnitude improvement in sensitivity relative to all previously-flown far-infrared observatories.** With a temperature of 4.5 K, *Origins'* sensitivity is limited by astronomical background photon noise (lower black curve). SOFIA (220 K), *Herschel* (80 K), and JWST (40 K) are shown for comparison with *Origins* (4.5 K). *Origins'* sensitivity extends JWST/MIRI sensitivity in mid-IR to the far-IR wavelengths.

**Unanticipated, yet transformative, discovery space:** The impressive *Origins*-enabled scientific advances discussed above are extensions of known phenomena. However, history has shown that order-of-magnitude leaps in sensitivity (Figure 8 and 9) lead to discoveries of unanticipated phenomena. For example, the sensitivity of IRAS over balloon and airborne infrared telescopes allowed the discovery of debris disks, protostars embedded within dark globules, Galactic infrared cirrus, and IR-bright galaxies, none of which were expected at the time of launch. Likewise, no study anticipated that *Spitzer* would study $z > 6$ galaxies, measure winds transporting energy in exoplanet atmospheres, and detect the dust around white dwarfs produced by shredded asteroids.

# 3. Technical Overview

## 3.1 Science Traceability

**Table 2 Science drives *Origins* key design parameters (Scientific Traceability Matrix (STM) provided in the full report)**

| *Origins* science driver | | Technical or instrument parameter | | | |
|---|---|---|---|---|---|
| Scientific goal | Observable | Parameter | Requirement | Design | Rationale |
| How do galaxies form stars, make metals, and grow their central SMBHs? | Mid- and far-IR rest-frame spectral lines. | Telescope Size | 3.0—5.0 m | 5.9 m | 5.0 m aperture is driven by the sensitivity to detect z > 6 galaxies; >3.0 m based on the sensitivity needed to detect z > 2 galaxies. |
| | | Telescope Temperature | < 6 K | 4.5 K | Sufficiently cold temperature to meet the sensitivity requirements at the longest wavelengths; $T_{tel}$ >6 K impacts spectral line sensitivity at $\lambda$ >350 $\mu$m. |
| How do the conditions for habitability develop during the process of planet formation? | $H_2{}^{18}O$ ground state 547.4 $\mu$m | $\lambda_{max}$ | > 550 $\mu$m | 588 $\mu$m | $H_2{}^{18}O$ ground state line and the need to measure continuum around it; $\lambda$ < 500 $\mu$m impacts extragalactic sciences. |
| | $H_2{}^{18}O$ 179.5-$\mu$m line | $R=\lambda/\Delta\lambda$ | 200,000 | 203,000 | Spectral resolving power is needed for Doppler tomography to connect water emission lines to disk location. |
| | HD 112-$\mu$m line | Spectral line sensitivity | $10^{-20}$ W $m^{-2}$ (1 hr; 5$\sigma$ ) | 5x$10^{-21}$ W $m^{-2}$ (1 hr; 5$\sigma$) | This sensitivity is required to measure disk gas masses and obtain a useful sample of the population of disks at the distance of Orion. |
| | | $R=\lambda/\Delta\lambda$ | 40,000 | 43,000 | The spectral resolving power needed for accurate gas mass measurements. |
| Do planets orbiting K- & M-dwarf stars support life? | $CH_4$ (3.3, 7.4 $\mu$m), $N_2O$ (4.5, 7.8 $\mu$m), $O_3$ (9.7 $\mu$m), $CO_2$ (4.3, 15 $\mu$m), $H_2{}^{18}O$ (6.3,17+ $\mu$m) | $\lambda_{min}$ | < 3 $\mu$m | 2.8 $\mu$m | $CO_2$ at 4.3 $\mu$m is the strongest of all features; $\lambda_{min}$ >5 $\mu$m reduces the exoplanet case to surface temperature only. |
| | | Aperture Size | 5.3 m | 5.9 m | Aperture size determines the sensitivity to detect faint $CH_4$ and $N_2O$ lines, crucial for biosignature detection in exoplanet transits over a 5-year mission. |

The three main science themes define the science traceability matrix (Table 2) for the Origins Space Telescope design.

***Origins* is >1000 times more sensitive than prior far-infrared missions and the design avoids complicated deployments to reduce mission risk.** The scientific objectives summarized in Table 1 are achievable with the low-risk *Origins* design. *Origins* has a *Spitzer*-like architecture (Figure 10) and requires only a few simple deployments to transform from launch to operational configuration. With the attributes shown in Table 3, the current design carries significant margin between science-driven measurement requirements and estimated performance (Table 2), leaving room for modest descopes.

**Table 3: *Origins* observatory-level parameters**

| **Mission Parameter** | **Value** |
|---|---|
| Telescope: Aperture Diameter/Area | 5.9 m/25 $m^2$ |
| Telescope Temperature | 4.5 K |
| Wavelength Coverage | 2.8—588 $\mu$m |
| Maximum Scanning Speed | 60″ per second |
| Mass: Dry/Wet (with margin) | 12000 kg/13000 kg |
| Power (with margin) | 4800 W |
| Launch Year | 2035 |
| Launch Vehicle | SLS Block IB or Space-X BFR |
| Orbit | Sun-Earth L2 |
| Propellant lifetime | 10 years, serviceable, limited by station-keeping propellant |

*Origins* provides a thousand-fold improvement in the far-infrared sensitivity relative to *Herschel* (Figure 8 and 9). While *Origins* has a 2.8x *Herschel*'s collecting area, cryocooling is the dominant factor affecting enabling its extraordinary sensitivity gain. To achieve the same sensitivity gain at optical wavelengths, the light-collecting area would have to increase a thousand-fold. A far-IR telescope limited in sensitivity by the astronomical background is essential to achieving the *Origins* science goals. Earth's warm atmosphere limits SOFIA's sensitivity and *Herschel* was limited by a relatively-warm telescope (70 K). The cryo-thermal system design of *Origins* leverages *Spitzer* experience and technology developed for JWST. Four current-state-of-the-art cryocoolers cool the telescope to 4.5 K, with 100% margin in heat-lift capacity at each stage. The science requirements can be met with a telescope temperature below 6 K.

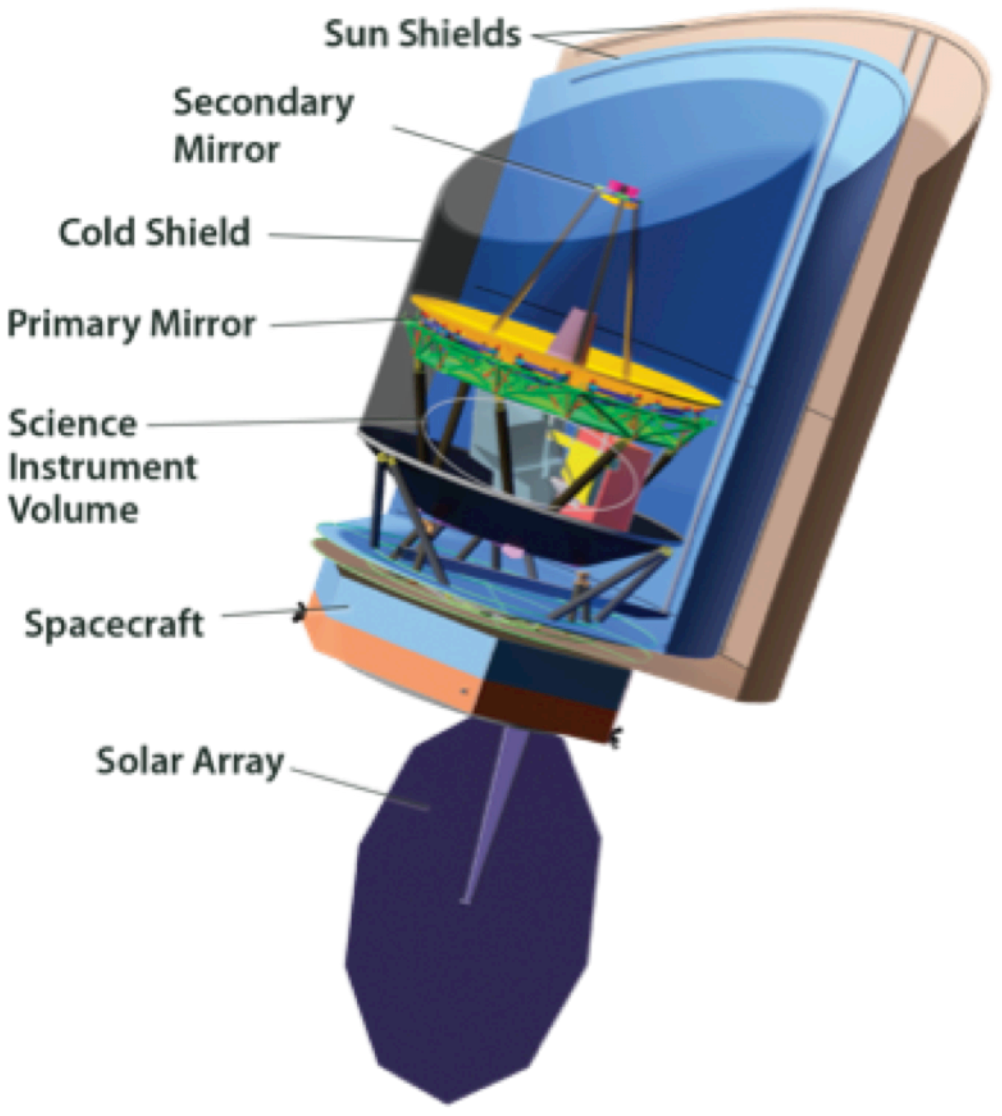

**Figure 10 : *Origins* builds on substantial heritage from *Spitzer* to minimize schedule risks during assembly, integration and testing, and deployment risks in space.** A cutaway view shows the locations of *Origins* instruments and major elements of the flight system. *Origins*, with an aperture diameter of 5.9 m and a suite of powerful instruments, operates with spectral resolving power from 3 to $3x10^5$ over the wavelength range from 2.8 to 588 μm. *Origins* has the agility to survey wide areas, the pointing stability required to observe transiting exoplanets, and operates with >80% observing efficiency, in line with the

The telescope is diffraction limited at 30 μm. All of the telescope's mirrors and mirror segments can be diamond turned and rough polished to the required precision in existing facilities. The JWST primary mirror segment actuator design is reused, to allow the *Origins* primary mirror segments to be adjusted in space in three degrees of freedom (tip, tilt, and piston), enabling final alignment during commissioning. The telescope is used as a light bucket at wavelengths between 2.8 and 20 μm to perform transit spectroscopy for exoplanet biosignatures since spatial resolution is not a technical driver for that scientific objective.

The *Origins* design minimizes complexity. The optical system launches in its operational configuration, requiring no mirror, barrel, or baffle deployments after launch, but the design allows for mirror segment alignment on orbit to optimize performance. The two-layer sunshield deployment is simple and low risk and other deployment mechanisms - communication antenna, solar array, telescope cover - have extensive heritage. This departure from the JWST deployment approach is enabled by the capabilities of new launch vehicles, which are expected to be fully operational in the mid-2030s. The design is compatible with at least two, and possibly three such launch vehicles. The fully-integrated cryogenic payload assembly comprising the telescope, instruments, and cold shield can be tested cryogenically in Chamber A at NASA's Johnson Space Center, following NASA's favored "test-like-you-fly" approach.

The next generation of launch vehicles, including NASA's SLS, SpaceX's BFR, and Blue Origin's 7-m New Glenn, have much larger payload fairings than the 5-m diameter ones available today, enabling the launch of a large-diameter telescope that does not need to be folded and deployed. *Origins* operates in a quasi-halo orbit around the Sun-Earth L2 point. The observatory is robotically serviceable, enabling future instrument upgrades and propellant replenishment to extend the mission life beyond the 5-year design lifetime.

## 3.2 Origins Instruments

**Table 4: Instrument capabilities summary**

| Instrument/ Observing Mode | Wavelength Coverage (μm) | Field of View | Spectral Resolving Power ($R=\lambda/\Delta\lambda$) | Saturation Limits | Representative sensitivity 5 σ in 1 hr |
|---|---|---|---|---|---|
| Baseline Instruments | | | | | |
| **Origins Survey Spectrometer (OSS)** | | | | | |
| Grating | 6 bands cover 25—588 simultaneously | 6 Slits: 2.7′ × 1.4″ to 14′ × 20″ | 300 | 5 Jy @ 128 μm | $3.7\times 10^{-21}$ W $m^{-2}$ @ 200 μm |
| High Resolution w/Fourier Transform Spectrometer | 25—588 Total range scanned by FTS | Slit: 20″ × 2.7 to 20″ × 20″ | 43,000 × [112 μm/λ] tunable w/FTS scan length | 5 Jy @ 128 μm | $7.4 \times 10^{-21}$ W $m^{-2}$ @ 200 μm |
| Ultra-High-resolution w/Fabry-Perot | 100—200 Select lines scanned | One beam: 6.7″ | 325,000 × [112 μm/λ] | 100 Jy @ 180 μm | $\sim 2.8 \times 10^{-19}$ W $m^{-2}$ @ 200 μm |
| **Far-infared Imager Polarimeter (FIP)** | | | | | |
| | 50 or 250 (selectable) | 50 μm: 3.6′×2.5′<br>250 μm: 13.5′×9′<br>(109×73 pixels) | 3.3 | 50 μm: 1 Jy<br>250 μm: 5 Jy | 50/250 μm: 0.9/2.5 μJy confusion limits<br>50/250 μm: 120 nJy/ 1.1 mJy |
| Survey mapping | 50 or 250 (selectable) | 60″ per second scan rate, with above FOVs | 3.3 | 50 μm: 1 Jy<br>250 μm: 5 Jy | Same as above, time to reach confusion limit:<br>50 μm: 1.9 hours<br>250 μm: 2 millisec |
| Polarimetry | 50 or 250 (selectable) | 50 μm: 3.6′×2.5′<br>250 μm: 13.5′×9′ | 3.3 | 50 μm: 10 Jy<br>250 μm: 10 Jy | 0.1% in linear polarization, ±1° in pol. Angle |
| **Mid-Infrared Spectrometer Camera Transit Spectrometer (MISC-T)** | | | | | |
| Ultra-Stable Spectroscopy | 2.8—20 in 3 simultaneous bands | 2.8-10.5 μm: 2.″5 radius<br>10.5-20 μm: 1.″7 radius | 2.8-10.5 μm: 50-100<br>10.5-20 μm: 165-295 | K~3.0 mag<br>30 Jy @ 3.3μm | Assume K~9.85 mag M-type star, R=50 SNR/sqrt(hr)>12,900 @ 3.3 μm in 60 transits with stability ~5 ppm < 10.5 μm, ~20 ppm > 10.5 μm |
| **Upscope Instruments** | | | | | |
| **HEterodyne Receiver for Origins (HERO)** | | | | | |
| Spectral line observations | 110 to 620 μm, dual-frequency, dual-polarization | 0.'5 x 0.'5 to 2' x 2' | up to $10^7$ | NA | $6.4 \times 10^{-21}$ at 480μm<br>$7.3 \times 10^{-20}$ at 130μm |

| Mid-Infrared Spectrometer Camera | | | | | |
|---|---|---|---|---|---|
| Imaging | 5 - 28 µm | 3'x3' | 5 - 10 | TBD | 6.4 x 10-21 at 480µm |
| Spectroscopy | 5 - 28 µm | 3' x (0."38 to 1."12) | 300 | 4Jy @ 5µm<br>8Jy @ 10µm<br>20Jy@ 20µm<br>50Jy@25µm<br>@ R=300 | 7.3 x 10-20 at 130µm |

Three baseline science instruments spanning the wavelength range 2.8 to 588 µm provide the powerful, new spectroscopic and imaging capabilities required to achieve the scientific objectives see Table.

OSS is a highly capable spectrometer that covers the entire 25 to 588 µm band at moderate (R~300), high (R~$4\times10^4$), and ultra-high (R~$2\times10^5$) spectral resolving power. OSS uses six gratings in parallel to take multi-beam spectra simultaneously across the 25 to 588 µm window through long slits. In this grating mode, OSS spatially and spectrally maps up to tens of square degrees of the sky providing 3-D data cubes. When needed, a Fourier transform interferometer and an etalon provide high and ultra-high spectral resolving power, respectively, in a single beam, with insertable elements that redirect the light path. The three OSS spectroscopy modes are packaged into one instrument. To meet its performance requires improved detector sensitivity and larger pixel format size.

FIP is a simple and robust instrument that provides imaging and polarimetric measurement capabilities at 50 and 250 µm. FIP utilizes *Origins'* fast mapping speed (up to 60″ per second) to map one to thousands of square degrees. FIP's images will be useful for telescope alignment and public relations. FIP's rapid mapping makes photometric variability studies possible for the first time. To meet its performance requires improved detector sensitivity and larger pixel format size.

MISC-T measures R~50 to 300 spectra in the 2.8 to 20 µm band with three subsystems that operate simultaneously. MISC-T has no moving parts and is designed to provide exquisite stability and precision (<5 ppm between 2.8 to 10 µm, <20 ppm 11 to 20 µm). The optics design uses densified pupil optics that mitigate for observatory jitter. The improved stability relies on a planned improvement in detector stability including calibration.

The *Origins* mission concept study team also developed two upscopes instruments, which enhance the mission's scientific capability: The Heterodyne Receiver for Origins (HERO) and the MISC Camera module. HERO provides nine-beam spectral measurements of selectable lines between 110 and 620 µm bands, up to very high spectral resolving power around $10^7$. The MISC Camera enables mid-infrared imaging and spectroscopy ($R$=300) between 5 and 28 µm. In addition there are upscopes for the existing instruments including expanded FOVs for OSS and FIP, and additional FIP bands (100 & 500 µm). Potential descopes include reducing the baseline instruments' modes and decreasing the aperture diameter, and would impact the observatory's science capabilities.

## 4. Key Technologies

**Detectors, ancillary detection system components and cryocoolers are the only *Origins* enabling technologies currently below Technology Readiness Level (TRL) 5. The *Origins* Technology Development Plan outlines a path leading to TRL 5 by Phase A start in 2025.**

At far-infrared wavelengths, reaching the fundamental sensitivity limits set by the astronomical background (Figure 8) requires a cold telescope equipped with sensitive detectors. The noise equivalent power (NEP) required for FIP imaging is $3 \times 10^{-19}$ W Hz$^{-1/2}$, whereas the NEP needed for OSS for $R$=300 spectroscopy is $3 \times 10^{-20}$ W Hz$^{-1/2}$. Transition edge sensor (TES) bolometers and kinetic inductance detectors (KIDs) both show great promise, and the plan is to mature both technologies to TRL 5 and then down-select to a single technology at the beginning of Phase A. While the noise requirements for MISC-T's mid-IR detectors are not particularly challenging, 5 ppm stability over several hours must be demonstrated to meet *Origins* requirement. The Technology Development Plan mitigates risk by recommending the parallel maturation of HgCdTe arrays, Si:As arrays, and TES bolometers (Figure 11).

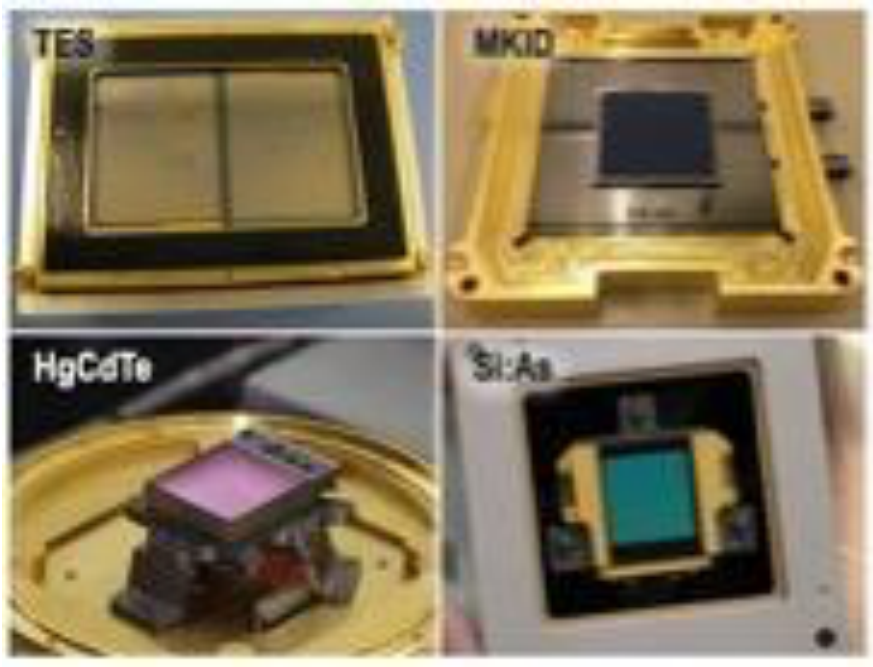

**Figure 11: The robust Origins Technology Development Plan recommends parallel maturation of multiple promising detector technologies (clock-wise from top-left):** TES bolometers (HAWC+ array); KIDS (432-pixel device); HgCdTe arrays (JWST/NIRCam); and Si:As (JWST/MIRI).

For the MISC upscope large format Si:As arrays and a deformable mirror capable of operation a ~8K are required. The increase in Si:As array format size is expected to be relatively straightforward as the JWST MIRI instrument already has detectors with the required performance, albeit a smaller array size. A cryogenic deformable mirror has been demonstrated, but at slightly higher temperatures.

The HERO upscope is based on the successful Herschel/HIFI receiver and is low risk. However, the HERO design uses the latest innovative components in order to substantially reduce weight, cooling power, and electrical power so that HERO can fly the first heterodyne array receivers on a satellite. Development of these components has already started, but needs to be continued to reach TRL 5 in 2025. The R&D includes broadband (hot electron bolometers and Superconducting Insulating Superconducting) mixers near the quantum limit; wideband local oscillators; low power, low noise cryogenic amplifiers; low power spectrometers and broadband optics.

Mechanical cryocoolers that can reach temperatures of 4.5 K have already flown on *Hitomi* (2016). These coolers, developed by Sumitomo Heavy Industries, had a required lifetime of 5 years compared to *Origins'* 10 years, but meet its performance requirements. Replacing the compressors' suspension system with a flex spring, a relatively straightforward change, will extend the lifetime. Several US companies have also produced TRL-5 cryocoolers or cryocooler components with a projected 10-year lifetime. The TRL 7 JWST/MIRI cryocooler, for example, has a 6 K operating temperature. Sub-Kelvin coolers operating at 50 mK, as needed for the OSS and FIP detectors, were also flown on *Hitomi*. A Continuous Adiabatic Demagnetization Refrigerator (CADR) with a much higher cooling power (6 μW vs. 0.4 μW for *Hitomi*), suitable for *Origins,* is currently being developed to TRL 6 under a Strategic Astrophysics Technology (SAT) grant. This new SAT CADR will also demonstrate self-shielding of magnetic fields to 1 μT, making it compatible with superconducting detectors that demand an ambient field of <30 μT. A straightforward extension of this ADR technology allows operations at even lower temperatures (35 mK), with similar cooling power. Lowering the operating temperature is a simple way to improve TES detector sensitivity, should that become necessary during mission formulation.

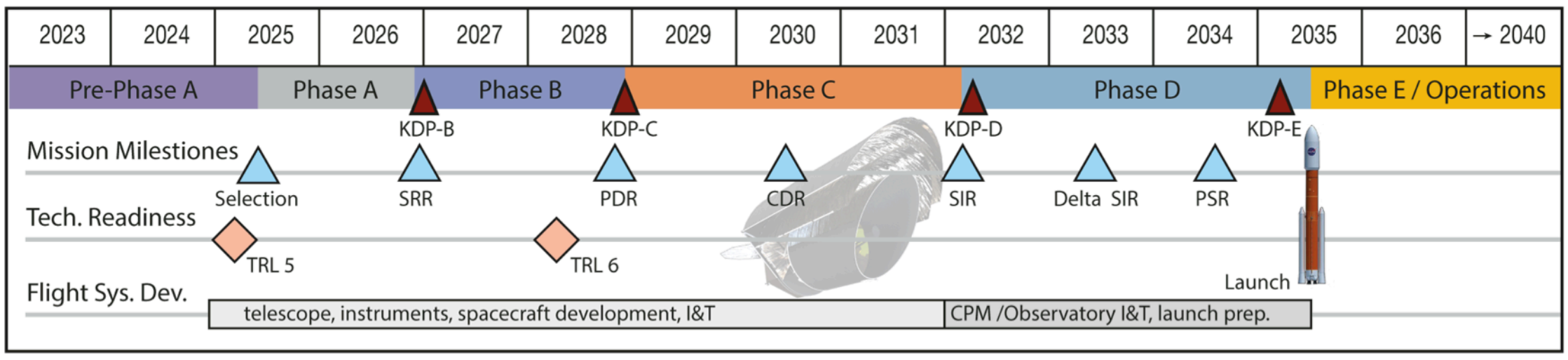

**Figure 12**: Schedule for Origins Space Telescope.

## 5. Schedule and Cost

**The *Origins* team developed a mission design concept, technical approach, technology maturation plan, risk management approach, budget, and a master schedule compatible with NASA guidelines for the Decadal Study and grounded in NASA and industry experience from previous successful large Class A missions.**

*Origins* is a NASA-led mission, managed by a NASA Center, and includes domestic and international partners. JAXA, and a CNES-led European consortium were active participants in the mission concept study, with each contributing an instrument design. US domestic participants included GSFC, Ames, MSFC, JPL, and the industry (Ball, Northrop, Lockheed, and Harris).

Figure 12 shows *Origins* Phase A through E schedule. Scheduled milestones and key decision points are consistent with formulation and development for Class A missions. The schedule supports an April 2035 launch, and includes 10 months of funded reserve. Much of the design and development work progresses through parallel efforts, with OSS as the critical path.

NASA Goddard Space Flight Center's Cost Estimating and Modeling Analysis (CEMA) office developed a cost estimate using the industry standard PRICE-H parametric cost modeling tool. The CEMA cost estimate is based on a detailed master equipment list (MEL) and the Integrated Master Schedule (IMS) shown in Figure 12 for the *Origins* baseline design. The MEL assigns an appropriate Technology Readiness Level (TRL) to each component. The CEMA cost model assumes that all components have matured to at least TRL 5 by the start of Phase A in 2025, and to at least TRL 6 by mission PDR. A separate *Origins Space Telescope* Technology Development Plan describes the maturation of all mission-enabling technologies on this timeline and reports the cost of technology maturation. The study team's mission cost estimate includes mission definition and development, the flight segment, the ground segment, and mission and science operations for 5 years. The launch cost ($500M for the SLS launch vehicle, as advised by NASA Headquarters) is also included. Working independently, Goddard's Resource Analysis Office (RAO) estimated the mission cost using a top-down parametric model. RAO and CEMA are firewalled from each other, but they both referred to the same MEL and mission schedule. The RAO and CEMA cost estimates agree to within 24%.

*Origins* is a "large" (>$1.5B) mission using the Decadal Survey's terminology. The NASA Headquarters-appointed Large Mission Concept Independent Assessment Team (LCIT) is tasked with validating the cost estimates supplied by each of the four large missions studied with NASA support. The study teams have decided to wait for feedback from the LCIT before publishing detailed cost information. The *Origins* Final Study Report will provide an LCIT-validated mission cost estimate.

The *Origins* mission design has not been optimized, and optimization may lead to cost savings. Optimization is planned as a Phase A activity. Japan and several ESA member nations have significant relevant expertise and have demonstrated interest in the *Origins* mission. Foreign contributions are expected to reduce NASA's share of the mission cost. NASA would welcome a European contribution equivalent in cost to an ESA M-class mission.

## 6. Possible European Contribution to Origins

| Table 5: Possible European Contribution | | |
|---|---|---|
| | **Possible Role** | **Description** |
| Mirror | lead | 5.9 m Silicon Carbide (SiC) mirror - Herschel heritage |
| OSS/FIPS Detectors | collaboration | Common development with US, European fabrication and delivery, US integration and testing |
| OSS/FIPS readout | lead | low power read-out electronics |
| OSS optics | collaboration | Etalon, FTS, grating, filters could be fabricated, tested and delivered by Europe |
| Sub-K cryocoolers | lead | 50 mK cooler for OSS/FIPS |
| HERO instrument | lead | All heterodyne components are available in Europe though a collaboration with the US is desirable |
| MISC instrument | collaboration with Japan and US | Components from Europe with JAXA taking the potential lead role. |
| FIP Instrument | collaboration or PI | far-IR detectors, cryogenic mechanisms |
| OSS Instrument | collaboration | Fourier-Transform Spectrometer scanning mechanism, the etalon, far-IR detector arrays, quasi-optical components |
| Concept Study | collaboration | |
| Science | collaboration | |
| Mission Operation & Data Analysis | collaboration | |
| Education/Outreach | collaboration | |

Table 5 lists some possible European contributions. NASA has welcomed international and industry participation during the Origins study and has stated its interest in mission partnerships to be discussed in phase A.

The segmented 5.9m mirror could be contributed by Europe. Europe has extensive experience in fabricating Silicon Carbide mirrors for space. The 3.5m mirror for Herschel and the 1.5x0.5m2 mirror for Gaia where fabricated in France and Germany.

The OSS and the FIPS instruments use detectors, readout and optics similar to that developed in Europe. A number of groups in the Netherlands and the United Kingdom are developing ultra-sensitive Kinetic Inductance Detectors and Transition-Edge Sensed (TES) bolometers, as well as the readout systems needed to operate these devices for SPICA/SAFARI and Athena/XIFU. The US design team has already exchanged with their European colleagues during the study. A potential partnership could be for one of the European groups to provide detector fabrication and delivery, after a common development period early in the project. A related possibility could be for the readout electronics, as French groups developed the readout for the Herschel/SPIRE and Planck/HFI bolometers, even though the detectors came from the US and were integrated into the instrument in the UK (SPIRE) and France (Planck/HFI).

Other possibilities include developing specific optical components. The etalon would be an excellent partnership opportunity since it is relatively modular and will require independent testing in advance of delivery. A group at SRON in the Netherlands is already developing new multi-layer mirrors that can make high-finesse etalons, and groups at Cardiff (UK) have experience with mirror materials and narrow-band etalon-based filters. These groups also have experience in Fourier Transform Spectrometers and are specialists in IR filters. They have delivered flight hardware, so would be well-positioned to contribute to OSS.

The sub-K cryogenics could be provided by French partners. The 300 mK systems for Herschel/SPIRE and /PACS and the 100 mK system for Planck came from France.

A largely European team designed the HERO instrument. Institutes in the Netherlands, Sweden, Germany, France, Italy and Spain are worldwide leaders in different components of heterodyne detector systems. They have contributed to the success of Herschel/HIFI as well as instruments on balloons, on SOFIA and for ground based telescopes.
The MISC instrument was led by a Japanese consortium and NASA/ARC, but France has made science and technical contributions. A collaboration between Japan, the US and Europe could enhance the instrument.

**Suggested European R&D in preparation for *Origins*:**
In order to maximize the scientific outcome of the mission and to allow for a valuable European involvement, feasibility studies and technical R&D in critical items should start as soon as possible.
As Europe (CNES) led the design of the Heterodyne REceiver for Origins (HERO), and as the European submillimeter community is very interested in spectroscopy, it offers itself as a potential European P.I. Instrument. The instrument is low risk as it has solid foundations thanks to the successful HIFI/Herschel instrument as well as smaller missions/projects. However, HERO would be the first heterodyne array receiver on a satellite and R&D work is required to increase its TRL (technology readiness level). A detailed technology roadmap has been developed and written up for HERO that can be provided upon request. The most critical developments are:

- *Mixers:* increasing sensitivity, stability and bandwidth of Hot Electron Bolometers
- *Local Oscillators*: Increasing RF bandwidth of Schottky amplifier-multiplier chains that are used as local oscillators;
- *Intermediate Frequency Chain*: Develop low-noise cryogenic amplifiers that consume very little power (< 0.5mW for 20dB gain);
- *Backends*: Develop spectrometers that consume very little power (< 2W, for 8 GHz bandwidth, 4 bits, 8k channels);
- *Optics:* Broadband, low-loss optics, in particular lenslet arrays and dichroic filters.

Cryogenic amplifiers, backends and optics are not only essential for HERO, but also part of the other far-IR instruments, FIP and OSS, on *Origins*.

The above-mentioned R&D work would not only greatly benefit *Origins*, but also other far-IR missions.

## 7. Origins in a worldwide context of space and ground telescopes

Origins builds on the heritage of Spitzer and Herschel, but surpasses them by a factor of 1000 in sensitivity in the mid to far-infrared wavelengths. Such a sensitivity gain allows an entirely new science discovery space with Origins that is impossible at UV/optical and X-ray wavelengths. The wish for a new mid/far-IR mission is shared by astronomers worldwide and there are two other proposed far-IR space missions: The Russian space agencies proposes the Millimetron Space Observatory (http://millimetron.ru/index.php/en/) with a 10m primary mirror cooled to about 35K. Millimetron is intended to harbor a comparable set of instruments, though these need to be contributed by Europe, Japan or the US. Should Millimetron launch as proposed before 2030 it could cover a large part of the Origins science. SPICA (https://spica-mission.org/) is another satellite similar to Origins, and is one of three candidates for ESAs M5 mission. SPICA is smaller (2.5m versus 5.9m primary), slightly warmer (8 versus 4.5K) and therefore a factor 100 less sensitive than Origins. Nevertheless it could do some preparatory science that Origins could follow up with larger statistics. SPICA does not have a stable mid-IR spectrometer for the search for biosignatures, nor have measurement capability with very high spectral resolving power in the far-IR (R~1000 versus R~200000/10000000 with HERO upscope), so most of the trail of water and spectral line observations remain unique to Origins. JWST will cover the near to mid-IR, but not extend to the far-IR. Nevertheless, JWST allows to do preparatory science, e.g. identification of exoplanets) and will complement Origins. There are also smaller missions planned that are pathfinders for Origins both for science and technological developments: E.g., the GUSTO balloon will carry small arrays of heterodyne receivers at 1.4, 1.9 and 4.7 THz to observe the major cooling lines with a 0.8m telescope. The SOFIA airplane observatory is equipped with far and mid-IR instruments and is an excellent testbed for technology as well as to advance infrared astronomy. Ground-based interferometers such as ALMA, SMA or Plateau de Bure interferometers, as well as single dish telescopes such as APEX, ASTE, CCAT prime and AtLAST complement Origins, as they allow observations at longer wavelengths and at higher spatial resolution. However, due to the atmospheric transmission they cannot operate in the mid or far-IR, which is unique to space.

### Origins Space Telescope Further Information

A full report of the study of the Origins Space Telescope will become available on line once it is submitted to the US Decadal later in 2019.

Origins website:
https://origins.ipac.caltech.edu/

Origins study center document repository:
https://asd.gsfc.nasa.gov/firs/

Origins Space Telescope Final Study Report and Technology Development Plan will be submitted to NASA HQ on August 23, 2019.

## Acknowledgements

The study was largely funded by National Aeronautics and Space Administration (NASA), which allowed us to hire engineering staff, subcontract parts of the study to industry and to pay for travel and meetings. The non-American team members are grateful for funding from their respective space agency and/or their research institute, in particular Centre National de Recherche Scientifiques (CNES), Swedish Natinal Space Agency, Netherlands Institute for Space Research (SRON), Japan Aerospace Exploration Agency (JAXA).

A large team of people carried out the Origins Space Telescope study. **The Science and Technology Definitions Team** (consisting of 25 American voting members, and 14 international non-voting members) defined the satellite in open meetings with additional interested scientists joining in the discussions, and the **NASA/GSFC Study Center Engineering Design team** carried out a preliminary design. There were also **US industry** studies and advice, as well as a **Study Advisory Board**. **Margaret Meixner** and **Asantha Cooray** are the joint P.I.s of the study.(Martina Wiedner is the instrument lead of the mostly European HERO study and design.)

**Origins Space Telescope (Origins) Science and Technology Definition Team (STDT) members**

Margaret Meixner, STScI/JHU, Baltimore, MD (Origins Study Chair)
Asantha Cooray, University of California at Irvine, Irvine, CA (Origins Study Chair)
Dave Leisawitz, NASA GSFC, Greenbelt, MD (Origins Study Scientist)
Johannes Staguhn, JHU/GSFC, Greenbelt, MD (Origins Deputy Study Scientist)
Lee Armus, California Institute of Technology/IPAC, Pasadena, CA
Cara Batterysb, University of Connecticut, Storrs, CT
James "Gerbs" Bauer, University of Maryland, College Park MD
Edwin "Ted" Bergin, University of Michigan, Ann Arbor, MI
Charles "Matt" Bradford, California Institute of Technology/JPL, Pasadena, CA
Kimberly Ennico-Smith, NASA Ames Research Center, Moffett Field, CA
Jonathan Fortney, University of California, Santa Cruz, CA
Tiffany Kataria, JPL/NASA, Pasadena, CA
Gary Melnick, Harvard Smithsonian, Boston, MA
Stefanie Milam, NASA GSFC, Greenbelt, MD
Desika Narayanan, University of Florida, Gainesville, FL
Deborah Padgett, JPL/NASA, Pasadena, CA
Klaus Pontoppidan, STScI, Baltimore, MD
Alexandra Pope, University of Mass, Amherst, MA
Thomas Roellig, NASA Ames Research Center, Moffett Field, CA
Karin Sandstrom, University of California at San Diego, in La Jolla, CA
Kevin Stevenson, STScI, Baltimore, MD
Kate Su, University of Arizona, Tucson, AZ
Joaquin Vieira, University of Illinois, Urbana-Champaign, Illinois
Edward L. "Ned" Wright, UCLA, Los Angeles, CA
Jonas Zmuidzinas, California Institute of Technology/JPL, Pasadena, CA

**Ex-Officio (non-voting) members of Origins Science and Technology Definition Team**

Kartik Sheth, NASA HQ, Program Scientist
Dominic Benford, NASA HQ, Deputy Program Scientist
Eric E. Mamajek, NASA JPL/Caltech, ExEP Deputy Program Scientist
Susan Neff, NASA/GSFC, COR Chief Scientist
Elvire De Beck, Chalmers Institute of Technology, SNSB Liaison
Maryvonne Gerin, LERMA, Obs. De Paris, CNES Liason
Frank Helmich, SRON, Netherlands Institute for Space Research Liaison
Itsuki Sakon, University of Tokyo, JAXA Liaison, MISC instrument lead
Douglas Scott, University of British Columbia, CAS Liaison
Roland Vavrek, ESA/ESAC, ESA Liaison
Martina Wiedner, LERMA, Obs. De Paris, HERO instrument lead
Sean Carey, Caltech/IPAC, Communications Scientist
Denis Burgarella, Laboratoire d'Astrophysique de Marseille, collaborator
Samuel Harvey Moseley, NASA/GSFC, instrument scientist

## References

Bolatto, A. et al. 2019, "Cold Gas Outflows, Feedback, and the Shaping of Galaxies," US Science White Paper, arxiv:1904.02120

Boogert, A. C. A., Gerakines, P. A., Whittet, D. C. B., 2015. "Observations of the icy universe," Annual Review of Astronomy and Astrophysics, 53, 541.

Caselli, P. et al., 2012. "First Detection of Water Vapor in a Pre-stellar Core," The Astrophysical Journal Letters, 759, 2, L37.

P. Hartogh, D. C. Lis, D. Bockelée-Morvan, M. de Val-Borro, N. Biver, M. Küppers, M. Emprechtinger, E. A. Bergin, J. Crovisier, M. Rengel, R. Moreno, S. Szutowicz, and G. A. Blake, "Ocean-like water in the Jupiter-family comet 103P/Hartley 2," *Nature*, **478**, 218–220, doi:10.1038/nature10519, 2011.

Shaposhnikov, D. S.; Medvedev, A. S.; Rodin, A. V.; Hartogh, P.: Seasonal Water ""Pump" in the Atmosphere of Mars: Vertical Transport to the Thermosphere." Geophysical Research Letters 46 (8), S. 4161 - 4169 (2019)

Kataria, T. et al. 2019, "The Mid-Infrared Search for Biosignatures on Temperate M-Dwarf Planets", US Science White Paper no 462

Keto, E., Rawlings, J., Caselli, P., 2014. "Chemistry and radiative transfer of water in cold, dense clouds," Monthly Notices of the Royal Astronomical Society, 440, 3, 2614.

Keto, E., Rawlings, J., Caselli, P., 2015. "The dynamics of collapsing cores and star formation," Monthly Notices of the Royal Astronomical Society, 446, 4, 3731.

M. Küppers, L. O'Rourke, D. Bockelee-Morvan, V. Zakharov, L. Seungwon, P. von Allmen, B. Carry, D. Teyssier, A. Marston, T. Müller, J. Crovisier, M.A. Barucci, R. Moreno, "Localized sources of water vapour on the dwarf planet (1)Ceres," Nature, 505, 525-527, doi: 10.1038/nature12918, 2014

P. Hartogh, E. Lellouch, R. Moreno, D. Bockelée-Morvan, N. Biver, T. Cassidy, M. Rengel, C. Jarchow, T. Cavalié, J. Crovisier, F. P. Helmich, and M. Kidger, "Direct detection of the Enceladus water torus with Herschel", Astron. & Astrophys., 532, L2, doi:10.1051/0004-6361/201117377, 2011.

Moreno, E. Lellouch, L. Lara, H. Feuchtgruber, M. Rengel, P. Hartogh, and R. Courtin, "The abundance, vertical distribution and origin of H2O in Titan's atmosphere: Herschel observations and photochemical modeling", Icarus, 221, 753–767, doi:10.1016/j.icarus.2012.09.006, 2012.

Öberg, K. I., Murray-Clay, R., Bergin, E. A., 2011. "The Effects of Snowlines on C/O in Planetary Atmospheres," The Astrophysical Journal Letters, 743, L16.

Pope, A. et al. 2019, "Simultaneous Measurements of Star Formation and Supermassive Black Hole Growth in Galaxies", US Decadal White Paper arxiv:1903.05110

Sadavoy, S. et al. 2019, “The Life Cycle of Dust,” BAAS, 51, 66

Smith, J. D. et al. 2019, “The Chemical Enrichment History of the Universe,” BAAS, 51, 400

L. Villanueva, M. J. Mumma, R. E. Novak, H. U. Käufl, P. Hartogh, T. Encrenaz, A. Tokunaga, A. Khayat, and M. D. Smith, “Strong water isotopic anomalies in the martian atmosphere: Probing current and ancient reservoirs”, Science, 348, aaa3630, doi:10.1126/science.aaa3630, 2015.

Whittet, D. C. B., et al. 1983. “Interstellar ice grains in the Taurus molecular clouds.” Nature. 303, 218.